# Heusler, Weyl, and Berry

Kaustuv Manna[1], Yan Sun[1], Lukas Müchler[1,2], Jürgen Kübler[3], Claudia Felser[1,*]

[1]*Max Planck Institute for Chemical Physics of Solids, 01187 Dresden, Germany.*
[2]*Department of Chemistry, Princeton University, Princeton, New Jersey 08544, USA.*
[3]*Institut für Festkörperphysik, Technische Universität Darmstadt, D-64289 Darmstadt, Germany*

**Abstract**

Heusler materials, initially discovered by Fritz Heusler more than a century ago, have grown into a family of more than 1000 compounds, synthesized from combinations of more than 40 elements. These materials show a wide range of properties, but new properties are constantly being found. Most recently, by incorporating heavy elements that can give rise to strong spin-orbit coupling (SOC), non-trivial topological phases of matter, such as topological insulators (TIs), have been discovered in Heusler materials. Moreover, the interplay of symmetry, SOC and magnetic structure allows for the realization of a wide variety of topological phases through Berry curvature design. Weyl points and nodal lines can be manipulated by various external perturbations, which results in exotic properties such as the chiral anomaly, and large anomalous spin and topological Hall effects. The combination of a non-collinear magnetic structure and Berry curvature gives rise a non-zero anomalous Hall effect, which was first observed in the antiferromagnets $Mn_3Sn$ and $Mn_3Ge$. Besides this *k*-space Berry curvature, Heusler compounds with non-collinear magnetic structures also possess real-space topological states in the form of magnetic antiskyrmions, which have not yet been observed in other materials. The possibility of directly manipulating the Berry curvature shows the importance of understanding both the electronic and magnetic structures of Heusler compounds. Together, with the new topological viewpoint and the high tunability, novel physical properties and phenomena await discovery in Heusler compounds.

**Introduction**

**Heusler** compounds were first discovered in 1903 by Fritz Heusler when he reported the surprising observation of room temperature ferromagnetic order in $Cu_2MnAl$, although none of its constituent elements, Cu, Mn, or Al, shows magnetism[1,2]. The crystal structure was unknown for quite a long time until 1934, when Heusler's son Otto Heusler[3] and Bradley[4] solved the crystal structure of $Cu_2MnAl$. Today, Heusler compounds represent a vast class of materials containing more than 1000 compounds with various fascinating properties important for many technological applications[5-11]. Heusler compounds host a plethora of unexpected exotic properties, which do not derive from only the properties of the atoms in the crystal structure.

**Weyl** fermions are the massless relativistic particles first proposed by Hermann Weyl in 1929 for a simplified model of the Dirac equation relevant in high-energy physics[12,13]. However, over so many decades, no such particle was ever detected in any high-energy experiments. Very recently and in the completely different context of condensed-matter physics, it was observed that the electronic excitations near the band crossings for certain semimetals such as TaAs[14-18] follow the same Hamiltonian that Weyl proposed in particle physics. The excitations close to the doubly degenerate Weyl point are described by a $2 \times 2$ Weyl equation, $H_w = \pm c\vec{p} \cdot \vec{\sigma}$, which can be viewed as half of the Dirac equation in 3D. Here $\vec{p}$ is the momentum operator and $\vec{\sigma}$ is the vector composed of the Pauli matrices. The sign, $k = \pm 1$, defines the chirality of the particle. Since a two-band model Hamiltonian can be generally described in the form of the linear combination of the three Pauli matrices, the linear band degeneracy can be obtained by tuning the parameters without any additional symmetries. In this case the expansion coefficients are three velocities and the chirality is generalized to $k = sign(\vec{v_1} \cdot \vec{v_2} \times \vec{v_3})$. Therefore, the only symmetry needed for the Weyl point is the lattice periodic boundary condition, and hence is much more robust against perturbations.

**Berry** reported in 1984 that the energy-level crossing of a Hamiltonian can lead to an object that behaves as a magnetic monopole, the Berry curvature[19]. Weyl points are such crossings and they behave as the sinks and sources of Berry curvature. The Berry curvature is the equivalent of the magnetic field in momentum space that illustrates the entanglement between the valence and conduction bands in the band structure. The integration of the Berry curvature on a closed sphere with one Weyl point inside yields the Chern number, which is the topological invariant of the Weyl point that characterizes it fully[20]. It is interchangeably referred to as a topological charge, and the sign of the Chern number defines the chirality of the Weyl point. The topological stability of Weyl semimetals (WSMs) depends on the separation of the pair of Weyl points and the only way to annihilate them is to move them to the same point in a Brillouin zone (BZ)[14,21-23]. An important consequence of Weyl points in the band structure is the formation of a non-closed Fermi arc on the surface and unusual transport properties such as the chiral anomaly[24,25] and a giant anomalous Hall effect (AHE)[26,27].

The AHE can have two origins, an extrinsic contribution from scattering and an intrinsic contribution from the Berry curvature of the electronic structure[27]. The intrinsic contribution is calculated by integrating the Berry curvature over the whole BZ[26,28,29]. Because the Berry curvature is odd under time reversal and Weyl points behave as the monopole of Berry curvature, WSMs that break time reversal symmetry are excellent candidates to observe a large intrinsic AHE. For an ideal WSM with two Weyl points located at $\pm k_w$, the anomalous Hall conductivity (AHC) is proportional to the distance between the pair of Weyl points[26,30-32]. From this point of view, magnetic WSMs can be viewed as close relatives of the quantum anomalous Hall system in 3D. In contrast to topological trivial magnetic metals, the density of state in WSMs can be extremely low and even vanishes at the Weyl points, leading to a very small charge carrier density, in addition to its large AHC. In the absence of other topological trivial bands, a large anomalous Hall angle (AHA) due to the low charge carrier concentration can be therefore be achieved in magnetic WSMs, which enhances the possibility of the realization of quantum AHE at high temperatures.

In Heusler compounds, the variety of accessible electronic and magnetic properties together with the existence of a topological band order generates highly tunable and versatile multifunctional properties such as topological superconductivity or spin-structured topological surface states etc[33-37]. For application in modern devices, it is therefore essential to understand the origin and stability of the various exotic properties of these novel topological states. In this review, we introduce the Berry curvature as one guiding principle to explain topological phenomena in Heusler compounds. This principle can further be used to explore other tunable material classes with novel properties, e.g., perovskites[38-40] and shandites[41-45]. Experimental signatures such as the formation of Fermi arcs between the two Weyl points, drumhead surface states, Chiral anomaly, and the intrinsic anomalous and spin Hall effect, as well as the topological Hall effect etc. are the direct consequences of the non-trivial topological states in Heuslers. Here we summarize simple rules for how one can tune various topological phases of Heusler compounds via Berry curvature engineering for next generation topo-spintronics applications.

**Box-1: Multiple faces of Heusler compounds**

The composition of a Heusler compound can be 1:1:1 (*half*-Heusler) or 2:1:1 (*full* or *inverse*-Heusler)[5]. They are generally denoted as *XYZ* or *X$_2$YZ*, where *X*, *Y* are transition metal elements, with *X* the most electropositive, and *Z* is a main-group element. The space group is (i) centrosymmetric $Fm\bar{3}m$ (225) with Cu$_2$MnAl (L2$_1$) as the prototype for *full* Heusler, and (ii) non-centrosymmetric $F\bar{4}3m$ with Li$_2$AgSb as the prototype for *inverse* Heusler and MgAgAs as prototype for *half*-Heusler. The general structure is cubic, but a structural deformation via compression or elongation along one of the cubic [100] axes generates a tetragonal lattice with space group *I*4/*mmm*; a similar deformation along the [111] direction results in a hexagonal structure. Because of the large magneto-crystalline anisotropy in tetragonal Heuslers, the preferred magnetization orientation can be tuned from in-plane to out-of-plane directions and sometimes stabilize in a complex non-collinear order. The various Heuslers can be metallic or semiconducting with a tunable band structure realizing novel states such as spin-gapless semiconductors, half-metals, and magnetic semiconductors, and are important for various potential technological applications[5,6,46-50].

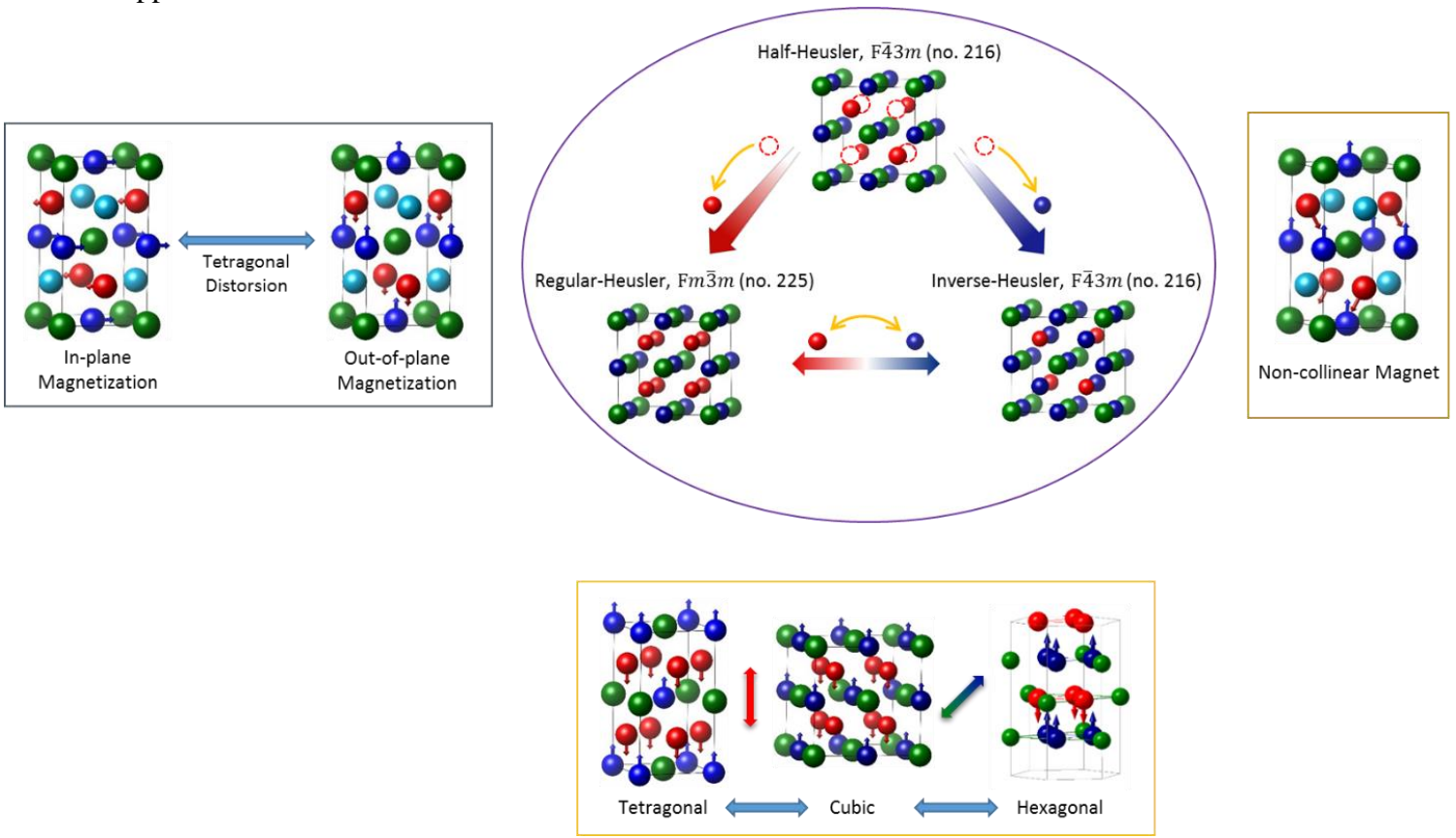

**Box2: Various topological states in Heusler compounds**

As a class of tunable materials, Heusler compounds host several topological phases, including topological insulators (TIs) and Weyl semimetals. One of the earliest discovered 3D TIs with a typical *s-p* band inversion (panel a) is a Heusler compound[51-55]. By extending topological band theory to semimetals, two types of topological semimetals were identified, where valence and conduction bands linearly touch each other resulting in Dirac and Weyl cones. By breaking time-reversal symmetry, Weyl semimetals can exist in ferromagnetic $Co_2MnAl$ and its counterparts (panel b), in which the location and numbers of Weyl points can be manipulated by the orientation of the magnetization direction[22,56-58]. In addition to *k*-space, Heusler compounds also host real-space topological states because of the existence of non-collinear magnetic structures. The absence of symmetries that reverse the sign of the Berry curvature in some non-collinear magnetic structures allows a non-zero AHE even at zero net magnetic moment, as was observed in the Heusler compounds $Mn_3Ge$ and $Mn_3Sn$ (panel c)[59-63]. A spin-Hall effect (SHE) is also proposed in these compounds (panel d)[64,65]. A strong *Dzyaloshinskii-Moriya (D-M)* interaction, in combination with $D_{2d}$ symmetry, lead to the observation of antiskyrmion structures in the tetragonal Heusler alloy Mn-Pt-Sn (panel e)[66].

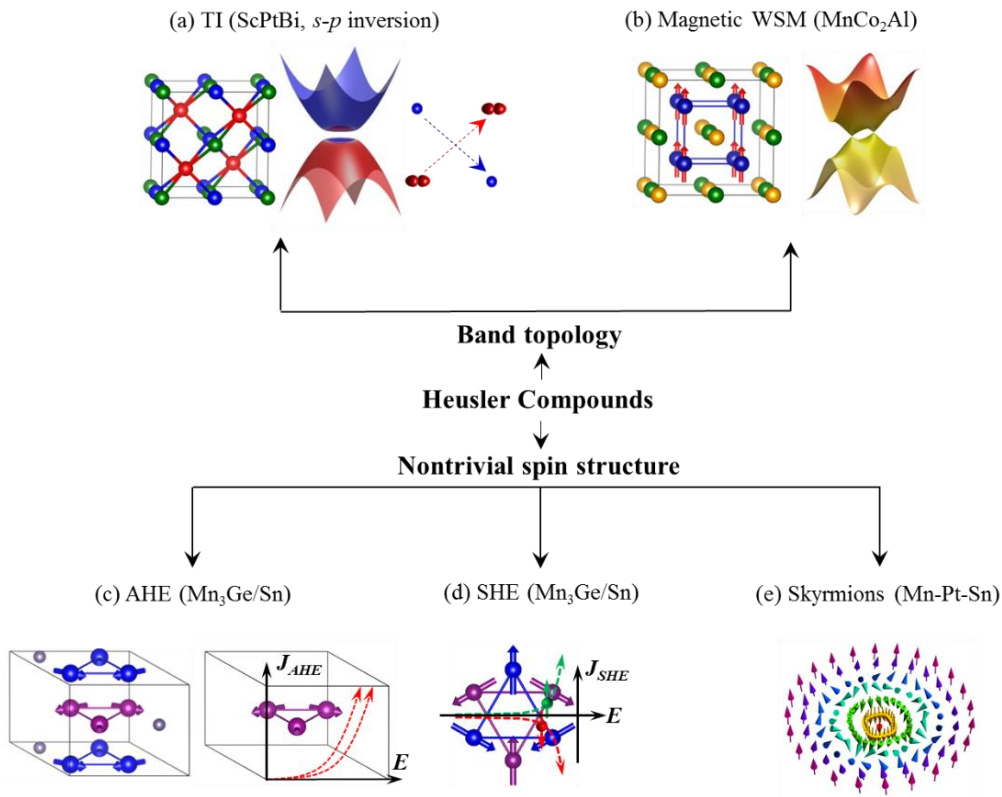

(a) TI (ScPtBi, *s-p* inversion)   (b) Magnetic WSM ($MnCo_2Al$)

**Band topology**

**Heusler Compounds**

**Nontrivial spin structure**

(c) AHE ($Mn_3Ge/Sn$)   (d) SHE ($Mn_3Ge/Sn$)   (e) Skyrmions (Mn-Pt-Sn)

# Weyl in *half*-Heusler:

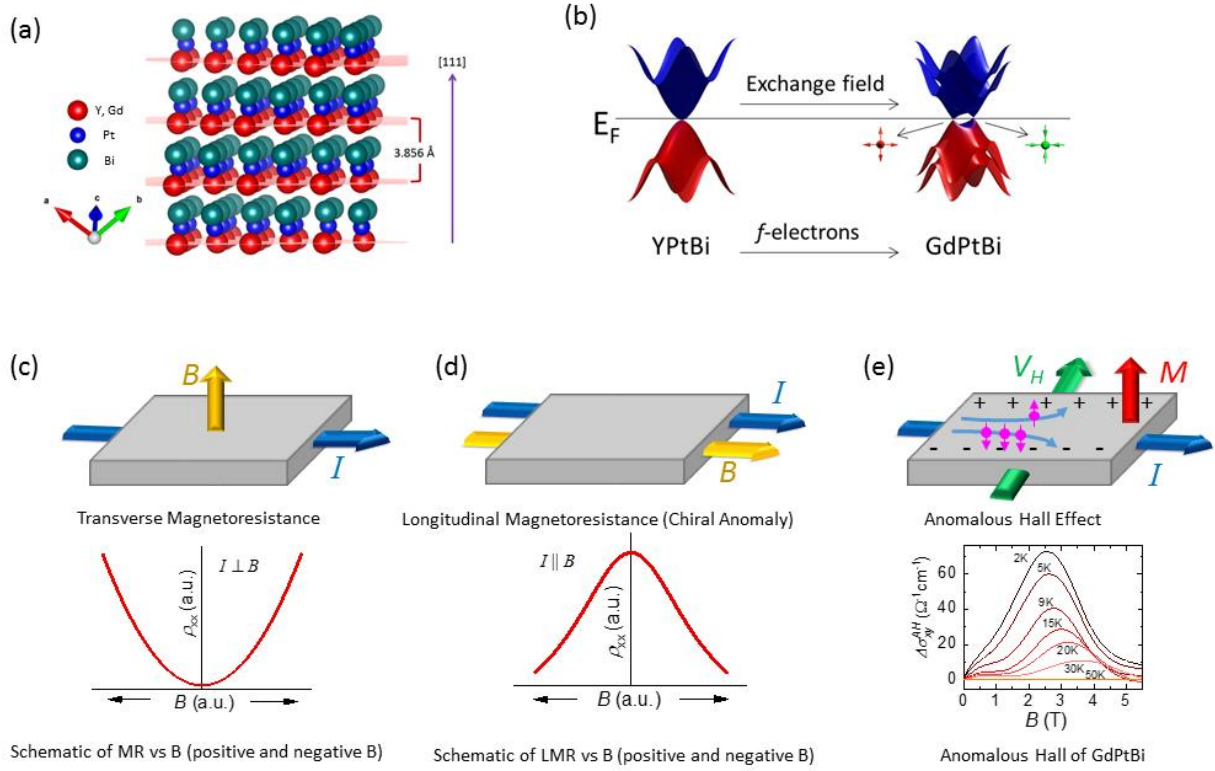

Fig. 1| **Weyl semimetal in GdPtBi.** (a) Crystal structure of *half*-Heusler GdPtBi. (b) Band structure evolution from quadratic touching to the Weyl point with applied external magnetic field. (c) Transverse magnetoresistance with $B \perp I$ and (d) chiral anomaly-induced negative magnetoresistance with $B // I$. (e) Intrinsic AHE in GdPtBi due to Weyl points.

Among the large variety of topological states found in the field of condensed-matter physics, the topological insulator (TI) is one of the most important. The first TI in HgTe/CdTe quantum wells were predicted by Bernevig *et al.*[51] in 2006 and experimentally verified by Koenig *et al.*[52] via the observation of a quantum spin Hall effect (SHE). In HgTe/CdTe quantum wells, the band inversion between the *s*-orbital-dominated $\Gamma_6$ state and the *p*-orbital-dominated $\Gamma_8$ state is the typical feature of the topological phase transition between normal and $Z_2$ TI. In 2010, similar electronic band structures were predicted in *half*-Heusler compounds by Chadov *et al.*[53], Lin *et al.*[54], and Di Xiao *et al*[55]. Similar to the binary zinc-blende semiconductors of HgTe and CdTe, the $s$-orbital-dominated $\Gamma_6$ state and $p$-orbital-dominated $\Gamma_8$ state also exist in a large number of *half*-Heusler compounds. Here the band gap and band order can be tuned by spin-orbit coupling (SOC), electronegativity difference of constituents, and lattice constants. In Heusler compounds, the trivial insulators show a positive band gap with $E_{\Gamma_6} - E_{\Gamma_8} > 0$ (such as ScPtSb). By increasing spin-orbital interaction (such as replacing Sb by Bi), a topological phase transition occurs along with *s-p* band inversion. A large family of tunable materials, there are more than 50 Heusler compounds predicted to have non-trivial band order[36,53,67], and some of them have been experimentally verified via magneto transport measurements or angle resolved photoemission spectroscopy (ARPES)[25,35,68,69].

The inverted band structure in Heusler compounds can be used to obtain a variety of other topological states; a typical example is the WSM. The *half*-Heusler compound GdPtBi (with Néel temperature $T_N = 9.2$ K) has an electronic structure with inverted band order and a quadratic band touching at the $\Gamma$ point[25,70,71]. However, the *f*-electrons from the Gd ions provide the possibility of tuning the electronic structure via control of the spin orientation. With Zeeman splitting, the spin-up and spin-down states near the Fermi level shift oppositely in energy, and Weyl points are formed between the shifted spin-polarized bands; see Fig. 1(a) and (b). The WSM state in GdPtBi was verified by the observation of different signatures of Weyl points, such as the chiral anomaly[25,70], unusual intrinsic AHE[25,72], nontrivial thermal effect[70], and strong planar Hall effect[73], as well as linear dependence of optical conductivity to temperature[74]. Owing to the defined chirality of each Weyl point, the charge carriers are pumped from one Weyl point to one with opposite chirality, when the magnetic field *B* is not perpendicular to electric field *E*. This breaks the conservation of Weyl fermions for a given chirality, which is the so-called chiral anomaly. The most important phenomenon induced by the chiral anomaly is the negative magnetoresistance (MR), which is shown in Fig. 1(d) for an ideal setup with *B // E*. As long as the magnetic field resides perpendicular to *E*, the negative MR disappears [Fig. 1(c)], implying that the contribution of negative MR originates only from the chiral anomaly of the Weyl points. Similar to electrical resistivity, Seebeck is thermal resistivity, where a thermal gradient is applied in place of the electrical gradient. In GdPtBi, variations similar to those of electrical resistivity are also observed in the thermal resistivity when a thermal gradient applied parallel to the applied magnetic field; this behaviour is also known as chiral anomaly because its origin is the same as that of Weyl[70]. Owing to the existence of Weyl points around the Fermi level, a sizable intrinsic AHC appears upon application of the magnetic field, see Fig. 1(e). Together with low charge carrier density and small longitudinal charge conductivity, the AHA can reach up to 10% in GdPtBi[25,72]. In addition to this chiral anomaly and AHC, GdPtBi exhibits an anomalously large value of 1.5 mΩcm planar Hall resistivity at 2 K in a 9 T magnetic field, which is completely different from the Hall resistivity[73]. Though the normal Hall signal is a function of the multiple of the sine and cosine of the applied field, a planar Hall signal is a function of cosine only. This is another alternative way to detect the chiral anomaly in Weyl semimetals.

**Half-metallic topological semimetal vs spin-gapless semiconductor**:

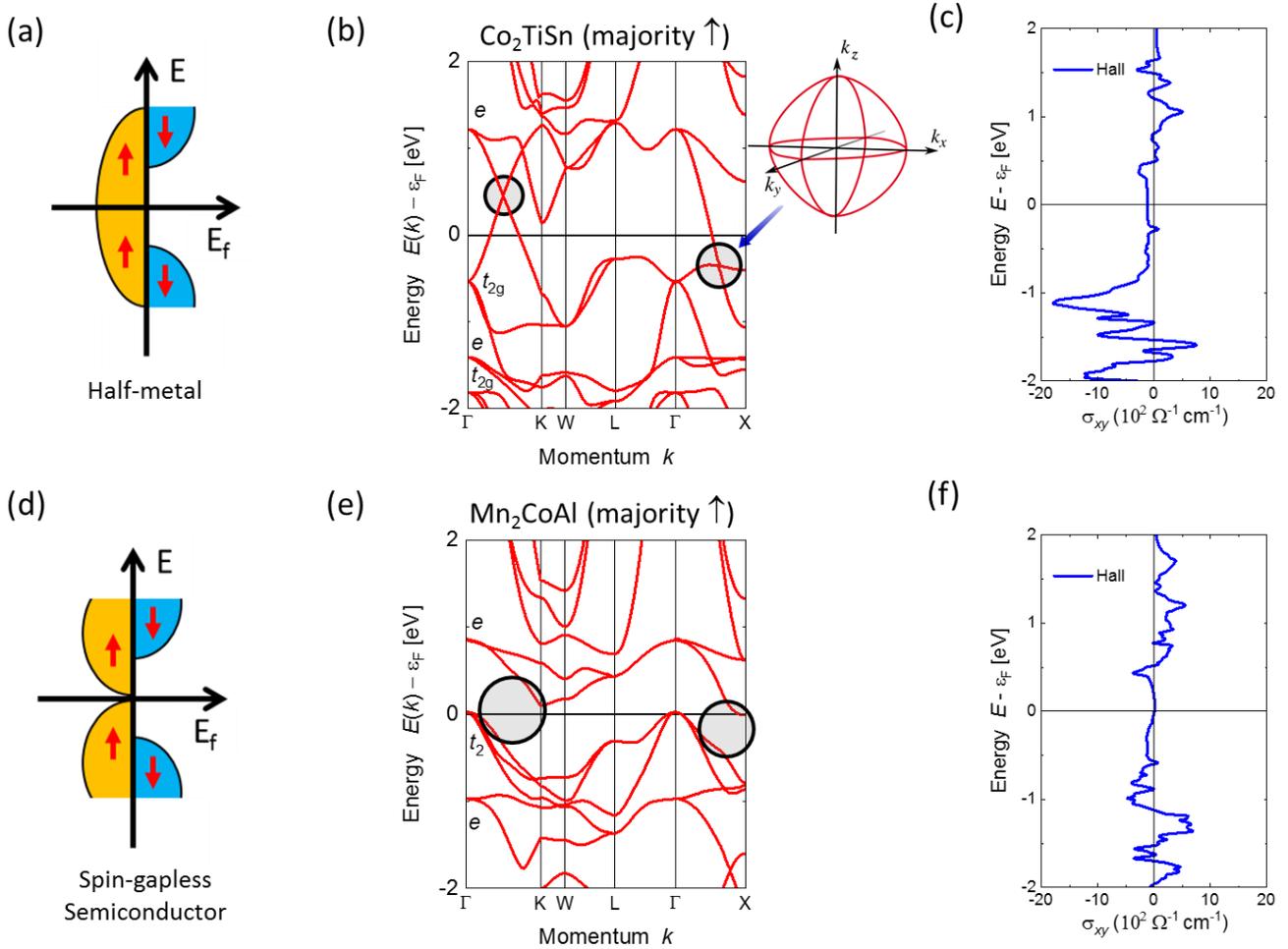

Fig. 2| **Influence of symmetry and Berry curvature on anomalous Hall effect.** Schematic density of states for (a) half-metallic ferromagnet and (d) spin-gapless semiconductor. The band structure and calculated energy-dependent anomalous Hall conductivity of (b, c) $Co_2TiSn$ and (e, f) $Mn_2CoAl$.

The concept of half-metallic ferromagnetism was first introduced by Groot *et al.*[75] in 1983; in it, one spin channel is insulating or semiconducting and the other spin channel is metallic because of ferromagnetic decoupling [Fig. 2(a)]. Because of the tunability of SOC, this half-metallic behavior plays an important role in the stability of topological semimetals in Heusler compounds, where the band crossings derive from bands with either the same or opposite spin-polarization [57,76]. Recently, topological surface states were predicted by Wang *et al.*[58] and Chang *et al.*[77] in half-metallic $Co_2$-based *full*-Heusler alloys. The half-metallic electronic structure goes in hand with several useful properties: (i) The spin orientation can be easily altered by a small external magnetic field because most half-metallic ferromagnets are soft-magnets (ii) The magnetic transition temperature is quite high, suitable for room temperature topo-spintronics applications[78-80]. (iii) Heusler compounds offer tunable band structures and symmetry elements by appropriate chemical substitution[6]. Therefore, the Berry curvature distribution can be easily changed and one can tune the anomalous Hall effect from zero to a very large value accordingly[26,81,82].

Considering $Co_2TiSn$ as an example, in Fig. 2(b), we find mirror symmetry protected band crossings between the valence and conduction bands close to $E_F$ as highlighted in the figure. When the bands of opposite eigenvalues cross, a nodal line is formed. Three such nodal lines form around the $\Gamma$ point in the $k_x$, $k_y$, and $k_z$ planes and are protected by the mirror symmetries $M_x, M_y, M_z$ of the $Fm\bar{3}m$ space group. Upon incorporating SOC, the electron spin is not a good quantum number any longer and the crystal symmetry changes depending on the direction of the magnetization. For example, if a sample is magnetized along the [001] direction, the $M_x$ and $M_y$ mirror symmetries are broken. Therefore, the nodal lines will open up unless there remains certain symmetries that protect the band crossings away from $E_F$. As a consequence, at least two Weyl points form along the $k_z$ axis, leading to a finite AHC. For example, we calculate an intrinsic anomalous Hall conductivity of ~ 120 $\Omega^{-1}cm^{-1}$ for the fully stoichiometric $Co_2TiSn$ compound. Depending on the details of the linear band crossings, such as the proximity of the nodal line to $E_F$ and its dispersion, the AHC in topological Heuslers can range from ~100 $\Omega^{-1}cm^{-1}$ in $Co_2TiSn$ to ~2000 $\Omega^{-1}cm^{-1}$ in $Co_2MnAl$[26]. The high AHE in $Co_2MnAl$ was already recognized in 2012 on the basis of Berry curvature calculation[26] and agrees well with the experiment[80]. However, the connection between the enhancement of Berry curvature and Weyl points or nodal lines was recognized later[56].

We now maintain the same number of valence electrons ($N_V$) and reduce the crystal symmetry in such a way that inversion and mirror symmetries are broken. An easy example is Mn$_2$CoAl with an *inverse*-Heusler structure $F\bar{4}3m$ that shares the same $N_V = 26$ of the *full*-Heusler compound Co$_2$TiSn. Interestingly, the compound belongs to a special class of materials, the spin-gapless semiconductor (SGS)[46]. Here, the minority spin channel is insulating, similar to the half-metallic compounds, but the majority spin channel possesses a vanishingly small gap at $E_F$ as shown in Fig. 2(d). Because of the non-centrosymmetric crystal structure, the mirror planes $M_x, M_y, M_z$ of the *full*-Heusler no longer exist. Naturally, the nodal lines gap, and upon incorporating SOC, no Weyl points form. Hence, the band structure of Mn$_2$CoAl does not show any topologically protected crossings as presented in Fig. 2(e). For the SGS compounds, the AHE shows an unusual behavior. Though the materials can be highly magnetic (saturation magnetization 2 $\mu_B$/f.u. for Mn$_2$CoAl), the AHC nearly compensates around $E_F$ [Fig. 2(f)], which is in contrast with the classical understanding that large magnetic moments always accompany a strong AHE. The predicted zero AHE was also found experimentally in Mn$_2$CoGa[82].

## Anomalous Hall effect in compensated ferrimagnets:

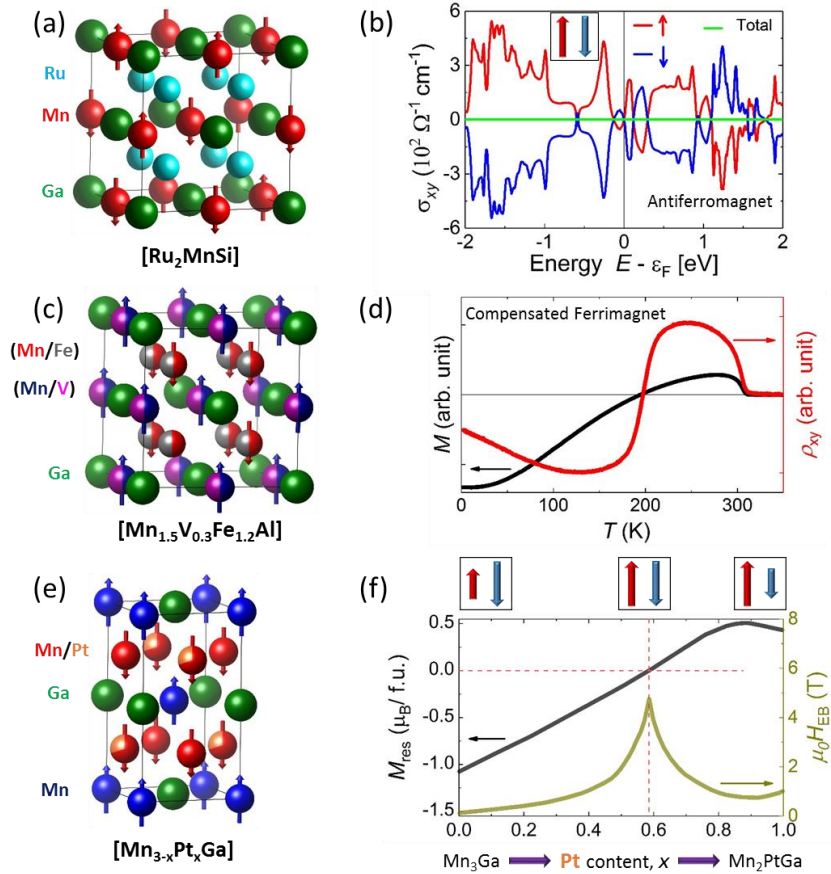

Fig. 3| **Effect of magnetic compensation on magnetization and transport.** (a) Crystal and spin configuration of antiferromagnetic $Ru_2MnSi$. (b) Energy dependence of the anomalous Hall effect of $Ru_2MnSi$. (c) Compensated ferrimagnetic structure of $Mn_{1.5}V_{0.3}Fe_{1.2}Al$ and (d) a comparative plot of temperature-dependent Hall resistivity and magnetization. (e) Tetragonal ferrimagnetic structure of $Mn_{3-x}Pt_xGa$. (f) Doping dependence of the residual magnetization and corresponding exchange bias.

Half-metallic materials are a natural source of spin-polarized current. Therefore, they are highly useful in spintronic devices as well as for spin manipulation applications[8,75,83]. However, ferromagnetic or ferrimagnetic compounds carry a major drawback. The net dipolar moment hinders the performance of nearby device arrays or the multilayer bits in spintronic chips. Certainly, materials with zero net magnetic moment are in high demand[84,85]. Half-metallic antiferromagnets that are stable in external magnetic fields are natural choices[86,87]. Unfortunately, the realization of such compounds remains a challenge, because antiferromagnetic (AFM) order implies an identical band structure of both spin channels.

Because of the excellent tunability of the Heusler compounds, the overall magnetic structure can be easily tuned via suitable chemical substitution in different magnetic sublattices. Therefore, the ground-state magnetic behavior can be interchanged from ferromagnetic to ferrimagnetic or to a completely compensated one. Because all these compounds generally follow the Slater-Pauling rule, the magnetic moment can be calculated using $M_S = N_V - 24$. Recently Stinshoff et al.[88-90] reported a half-metallic completely compensated ferrimagnetic state in the Heusler alloy $Mn_{1.5}V_{0.5}FeAl$ with a cubic $L2_1$ crystal structure. In fact, by a suitable tuning of the composition, the magnetic compensation temperature ($T_{MC}$) can be changed from *zero* to 226 K. The AHE of these compounds carry interesting features, which we compare to the characteristics of the pure AFM compound $Ru_2MnSi$.

The $L2_1$ crystal structure of $Ru_2MnSi$ is presented in Fig. 3(a). The compound possesses a so-called second kind of AFM structure, where the four constituent cubic sublattices follow a checkerboard-like AFM order[91]. The AHC contribution from the spin-up and spin-down channels are canceled out because of the combined time reversal and glide symmetry, resulting in a *zero* AHC [Fig. 3(b)] in the whole energy window at any measuring temperature. However, for the compensated AFMs, the net AHC depends on the details of the sublattice magnetization. We exemplify this by the isostructural compound $Mn_{1.5}V_{0.3}Fe_{1.2}Al$, which shows magnetic compensation of the sublattices at $T_{MC} \sim 127$ K[88,89]. The corresponding crystal structure and the spin distribution are illustrated in Fig. 3(c). The magnetization changes sign below $T_{MC}$ as the sample is cooled from the high-temperature paramagnetic state [Fig. 3(d)]. Evidently, the magnetic moment is *zero* at $T_{MC}$. For the other temperature regions, one sublattice magnetization dominates, and the sample's magnetization acquires the sign accordingly. The AHC of these magnetic

Heuslers scales with the sample's magnetization. Therefore, the Hall conductivity changes sign following the sublattice magnetization and acquires a *zero* value only at $T_{MC}$. However, in topological materials with special band structures, the AHE can behave differently. Because of the large Berry curvature around the Weyl point and low charge carrier density, a strong AHE and large AHA[92] is predicted in the compensated spin-gapless *inverse*-Heusler compound $Ti_2MnAl$. Owing to its *zero* net magnetic moment with $T_N \sim 650$ K[93], this can be viewed as an artificial AFM suitable for room temperature spintronics applications.

In contrast to the cubic Heuslers, the tetragonal Heusler compounds do not show half-metallicity, but still has a high spin polarization due to a pseudogap in one spin direction[49,94]. However, because of the large magnetorystalline anisotropy and high Curie temperature, the tetragonal Heuslers serve an important role as permanent magnets and in spin transfer torque related spintronic applications[94-98]. Mn-based compounds are generally ferrimagnetic in nature and crystalize in the *I*4/*mmm* space group[99,100]. Here the Mn atoms reside in two different crystallographic positions with different localized magnetic moments. By suitable chemical substitution, a completely compensated ferrimagnetic state can be easily obtained, such as in $Mn_{3-x}Y_xGa$ (where *Y* can be Ni, Cu, Rh, Pd, Ag, Ir, Pt, or Au)[101]. In the tetragonal $Mn_{3-x}Pt_xGa$ compound, the two symmetry-inequivalent Mn atoms possess different magnetic moments (3.1 $\mu_B$ compared to 2.1 $\mu_B$) [Fig. 3(e)]. Depending on the concentration of the non-magnetic atom Pt, the net magnetization changes from ferrimagnetic to a completely compensated one for $x \sim 0.6$ [Fig. 3(f)]. In this compound, the microscopic segregation of defect ferromagnetic phases influence the physical properties on a macroscopic level. An extremely large exchange-bias up to several Tesla is reported in these compounds with maximum effect around the magnetic compensation composition [Fig. 3(f)][99].

# Hall effect in non-collinear antiferromagnets:

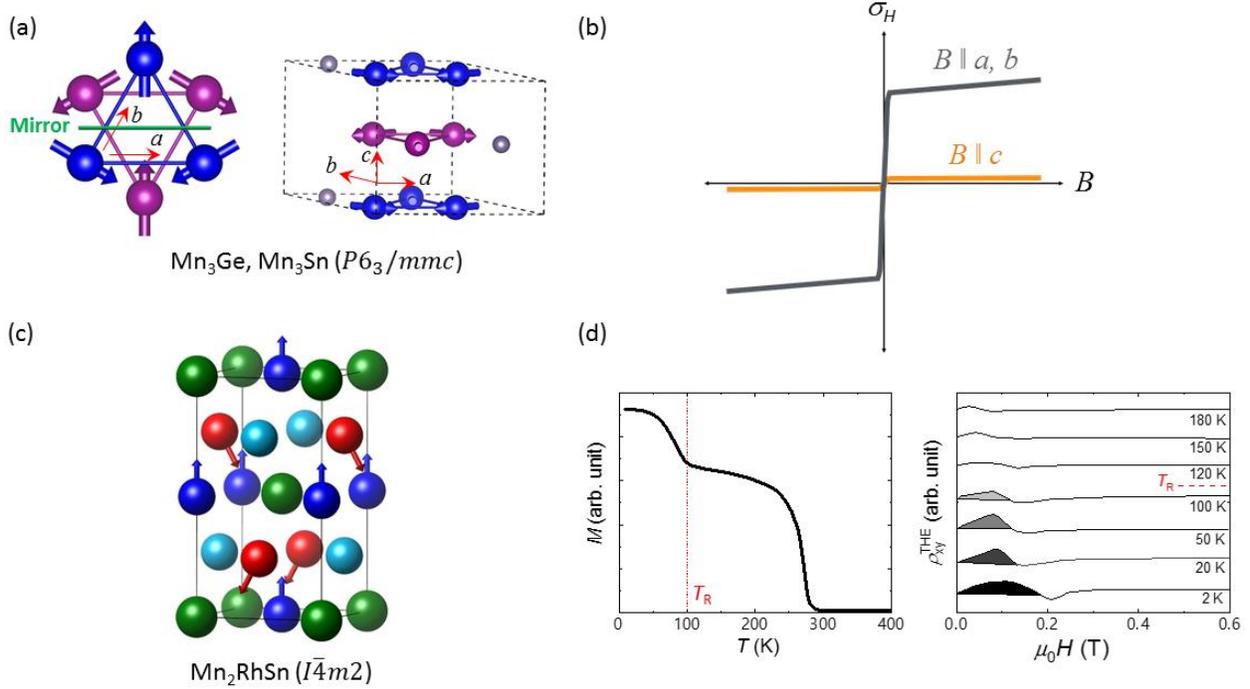

Fig. 4| **Non-collinear spin structure to tunable skyrmions.** (a) Hexagonal lattice structure and triangular magnetics in Mn₃Ge and Mn₃Sn. The green line represents the glide mirror plane $[M_y|\left(0,0,\frac{c}{2}\right)]$. (b) Schematic of the AHC when a magnetic field is applied along and perpendicular to *c*. (c) Crystal and non-collinear spin structure of Mn₂RhSn. (d) The magnetization and topological Hall effect in Mn₂RhSn.

### (i) Anomalous and spin Hall effects:

For a long time, it was believed that an AHE cannot exist in AFM materials due to the *zero net magnetic moment*. However, it was recently revealed that the existence of the AHE relies only on the symmetry of the magnetic structure and corresponding Berry curvature distribution. Since the AHC can be understood as the integral of Berry curvature in *k*-space, and the Berry curvature is odd under time-reversal operation, the AHE can exist only in systems with broken time-reversal symmetry. In collinear AFMs, the combined symmetry of time reversal $\hat{T}$ and a space group operation $\hat{O}$ will change the sign of Berry curvature ($\Omega_i(\vec{k}) = -\Omega_i(\hat{T}\hat{O}\vec{k})$), leading to a vanishing

AHC, despite the broken time-reversal symmetry due to the formation of local magnetic moments. However, in certain non-collinear AFMs, the symmetry to reverse the sign of the Berry curvature is absent, and a non-zero AHE can appear.

A non-collinear AFM order was first demonstrated by and Kren et al.[102] in 1968 for the cubic compounds $Mn_3Rh$ and $Mn_3Pt$. A similar spin structure in the hexagonal series of $Mn_3X$ ($X$ = Ga, Ge, Sn) compounds was discovered by Kren et al.[103], Nagamiya[104], Tomiyoshi et al.[105], and Brown et al[106]. These early investigations invoked the DM interaction[107,108] to explain the observed triangular order by neutron diffraction experiments. In 1988, the first ab initio density functional calculations were reported by Kübler et al.[109] for the $Cu_3Au$ structures of $Mn_3Rh$ and $Mn_3Pt$; they succeeded in explaining the observed non-collinear order. By the same method, Sticht et al.[110] dealt with the hexagonal $Mn_3Sn$, successfully obtaining and analysing the triangular magnetic ground-state structure. Later, Sandratskii et al.[111] showed that the DM interaction produces a weak ferromagnetism in $Mn_3Sn$. The DM vector is oriented along the crystallographic $c$-axis and leads to a negative chirality of the spin structure.

The first non-zero AHE in non-collinear AFMs was predicted in cubic $Mn_3Ir$ by Chen et al.[112] However, its experimental realization has not yet been successful. Motivated by theoretical studies of the stability of cubic, tetragonal, and hexagonal phases of $Mn_3X$ ($X$ = Ga, Sn, Ge) in connection with the Heusler family, a new series of studies began[113], which led to the prediction of the AHE in hexagonal $Mn_3Sn$ and $Mn_3Ge$ by Kübler et al.[59] Soon after the predictions, the large AHE were experimentally verified in both $Mn_3Sn$ and $Mn_3Ge$ hexagonal antiferromagnets [60,114,115].

The AHC can be viewed as a vector in three dimensions, where the non-zero components are determined by symmetry. Both $Mn_3Ge$ and $Mn_3Sn$ exhibit a triangular antiferromagnetic structure with an ordering temperature above 365 K, and the magnetic structure is symmetric with respect to the glide mirror operation $[M_y|(0,0,\frac{c}{2})]$ [left panel in Fig. 4(a)][60,114]. Under this symmetry operation, the two components of Berry curvature $\Omega_x$ and $\Omega_z$ change sign, whereas $\Omega_y$ does not. As a consequence, $\sigma_x$ and $\sigma_z$ are forced to be *zero*, and only a non-*zero* $\sigma_y$ survives. Therefore, a non-zero AHE can be obtained only when the magnetic field is applied perpendicular to *c* [Fig. 4(b)]. The maximum AHC appears for the setup with $B\perp a$ ($B//y$). Since there is a weak net moment (~ 0.01 $\mu_B$/Mn) out of the *a-b* plane, a very small AHC was also detected in the situation with $B//c$, which is orders of magnitude smaller than that of the other configuration[64,116].

The SHE is another important member in the family of Hall effects[27]. Owing to the net spin current in the absence of magnetic moments, SHE has been viewed as an important phenomenon for potential application in spintronics such as magnetoresistive random access memory (MRAM). Here, an electric current is passed through a sample leading to a pure spin current, which is polarized perpendicular to the plane defined by the charge and spin current (panel d of Box 2). It is generally induced by SOC, and a strong SHE can normally be observed in compounds with heavy transition metals[117]. In non-magnetic materials the mechanism is explained by the asymmetric Mott scattering[118]. In magnetic materials, the intrinsic SHE is explained by the spin Berry curvature that is obtained from a Kubo formula, similar to the AHE[64,117].

However, in non-collinear antiferromagnets, the understanding of SHE in the context of the spin currents had to be refined. Because of the special symmetry of chiral spin structures, a strong SHE can exist in non-collinear antiferromagnets, such as $Mn_3Ga$, $Mn_3Ge$ and $Mn_3Sn$[61,65]. This understanding provides a new guideline in the search for new materials with large SHE based on the light elements. Very recently, a strong SHE was observed in chiral antiferromagnetic $Mn_3Ir$, and the spin Hall angle can take on values up to ~35%[50]. Apart from the anomalous Hall and spin Hall effects, non-collinear antiferromagnets host another interesting phenomenon, i.e., the charge current is spin-polarized[116]. Similar to the AHE, in collinear antiferromagnets the spin-polarized current is not allowed by the symmetry. This makes non-collinear antiferromagnets uniquely suited for spintronic applications.

**(ii) Thermal Hall effect:**

The dissipation-less transverse charge flow due to the Berry curvature also produces thermal signals, such as the off-diagonal thermal effect (Righi-Leduc) and the thermoelectric effect (Nernst)[119]. Both effects have quite recently been explored for hexagonal $Mn_3Sn$[120]. Nernst- or Righi-Leduc-effects follow precisely the curve sketched in (b), provided we relabel the $\sigma_H$-axis by the Righi-Leduc coefficient $\kappa_{xz}$ or the Nernst coefficient $S_{xz}$ or the temperature change of the Peltier effect. The results of Ikhlas *et al.*[121] are similar, but the signals show hysteresis.

The important message from the work of Li *et al.*[120] is that the Wiedemann-Franz law is robust allowing us to *exclude any contributions from inelastic scattering*. Thus, the anomalous electric and thermal Hall currents are carried by the quasiparticles at the Fermi surface[28]. In the case of the AHE of $Mn_3Sn$ (and most likely also of $Mn_3Ge$) inelastic scattering plays no role. As explained previously, materials with a large AHE usually possess electronic band structures that are topologically nontrivial, and in the case of $Mn_3Sn$ and $Mn_3Ge$ it could be shown that Weyl points contribute largely to the AHE[62,122]. The interplay between anomalous transport effects and topological order is another interesting avenue to discover new topological states of matter.

**(iii) Antiskyrmions and topological Hall effect:**

Apart from a non-zero AHE, Heusler compounds with non-collinear spin structure also host real-space topological states such as magnetic skyrmions. Magnetic skyrmions are particle-like vortex spin textures surrounded by chiral boundaries that are separated from a region of reversed magnetization found in magnetic materials[66,123-125]. In this case, the topological skyrmion number is defined in real space. It measures the winding of the magnetization direction wrapped around the unit sphere and can take on integer values only. The mechanism of formation and stabilization of skyrmions can be understood as due to the competition of the ferromagnetic exchange and the relativistic DM interaction in non-centrosymmetric magnets. The typical size of a skyrmion can range from 1 to 100 nm, which enables the manipulation of many internal degrees of freedom. Owing to the magnetoelectric coupling, it is possible to control the skyrmions with an external electric field with low energy consumption.

Depending on the spin rotation, skyrmions can be classified into two fundamental types, Bloch skyrmions and Néel skyrmions[123]. Another type of skyrmion (antiskyrmion) was also proposed to exist, where the boundary domain walls alternate between the Bloch and Néel type as one traces around the boundary[66,126,127]. The first two fundamental types of skyrmions were observed in B20 crystals and polar magnets with $C_{nv}$ symmetry, respectively. However, despite a prediction of antiskyrmions in Co/Pt multilayers and B20 compounds, none have been experimentally verified. Very recently, by following the theoretical prediction and symmetry analysis, the first class of antiskyrmions has been observed in the inverse tetragonal acentric Mn-Pt-Sn Heusler compounds with D2d symmetry[66]. Here we summarize the various types of skyrmions reported in different compounds in Table- 1.

Table 1: Summary of the various types of skyrmions observed in different compounds, reported so far in literature.

| Allowed Point Group Symmetry | Type of Skyrmion | Compounds |
| --- | --- | --- |
| $T$ | Bloch | FeGe[128], MnSi[129], MnGe[130], Fe$_{0.5}$Co$_{0.5}$Si[131], Cu$_2$OSeO$_3$[132] |
| $O$ | Bloch | Co$_{10}$Zn$_{10}$[133], Co$_{10-x}$Zn$_{10-x}$Mn$_{2x}$[134,135] |
| $C_n$v ($n \geq 3$) | Néel | GaV$_4$S$_8$[136], VOSe$_2$O$_5$[137] |
| $D_{2d}$ | antiskyrmion | Mn$_{1.4}$PtSn[66], Mn$_{1.4}$Pt$_{0.9}$Pd$_{0.1}$Sn[66] |
| $D_{2d}$ | Predicted antiskyrmions | Mn$_2$RhSn[138,139], CuFeS$_2$[127,140], Cu$_2$FeSnSe$_4$[127,141], Cu$_2$Mn$_{1-x}$Co$_x$SnS$_4$[127,142] |

Because of the large magentocrystalline anisotropy of the tetragonal Heusler compounds, the preferred spin orientation can be switched from in-plane to out-of-plane by a suitable tetragonal distortion. Therefore, it is always possible to arrive at a situation where the anisotropy energy is such that the spin structure stabilizes in a non-collinear fashion. Interestingly, in many such compounds with D$_{2d}$ crystal symmetry, antiskyrmion phases are proposed to form. Due to the strong DM interaction, the localized spins of the Mn atoms in the two crystallographic positions twist in such a way that a helimagnetic structure is obtained. However, due to the constraints of the D$_{2d}$ crystal symmetry, the ferromagnetic order is restricted only in the tetragonal basal plane. Now, because of the superimposed effect of the strong magentocrystalline anisotropy, the antiskyrmionic states stabilize when a magnetic field is applied along the crystallographic $c$ axis[66]. These spin structures of skyrmions can be directly detected by various techniques such as small angle neutron scattering (SANS), Lorentz transmission electron microscopy (LTEM), and magnetic force microscopy (MFM). Furthermore, the existence of skyrmions can also be identified from the distinct transport signals.

As discussed previously, the Hall effect is one of the most powerful tools to probe various exotic physical characteristics of materials. Usually the Hall resistivity is expressed as $\rho_{xy} =$ $R_0\mu_0 H + R_S M + \rho^{Topo}$, where the first term is the linear Hall contribution due to the Lorentz force, the second term is the magnetic contribution, and the last term is the topological Hall effect (THE) due to the real-space Berry curvature, which originates from factors such as skyrmion formation etc. The THE has been viewed as one of the most important features of skyrmion formation, and widely used for the confirmation of skyrmions in transport experiments[138,143-145]. Fig. 4(c) presents the non-collinear spin structure of the tetragonal Heusler compound $Mn_2RhSn$. The compound shows a ferrimagnetic transition temperature of ~ 270 K, and upon cooling undergoes a spin-reorientation transition ($T_R$) below 100 K, resulting in a non-collinear spin structure[139,146]. Transport measurements on $Mn_2RhSn$ thin-films show a large THE below the spin-reorientation transition at 100 K [Fig. 4(d)][138]. Motivated by this large THE in the parent compound $Mn_2RhSn$, Nayak *et al*. performed a series of chemical substitutions and discovered the antiskyrmionic state in tetragonal Heusler Mn-Pt-Sn by LTEM[66].

## Conclusions and future direction:

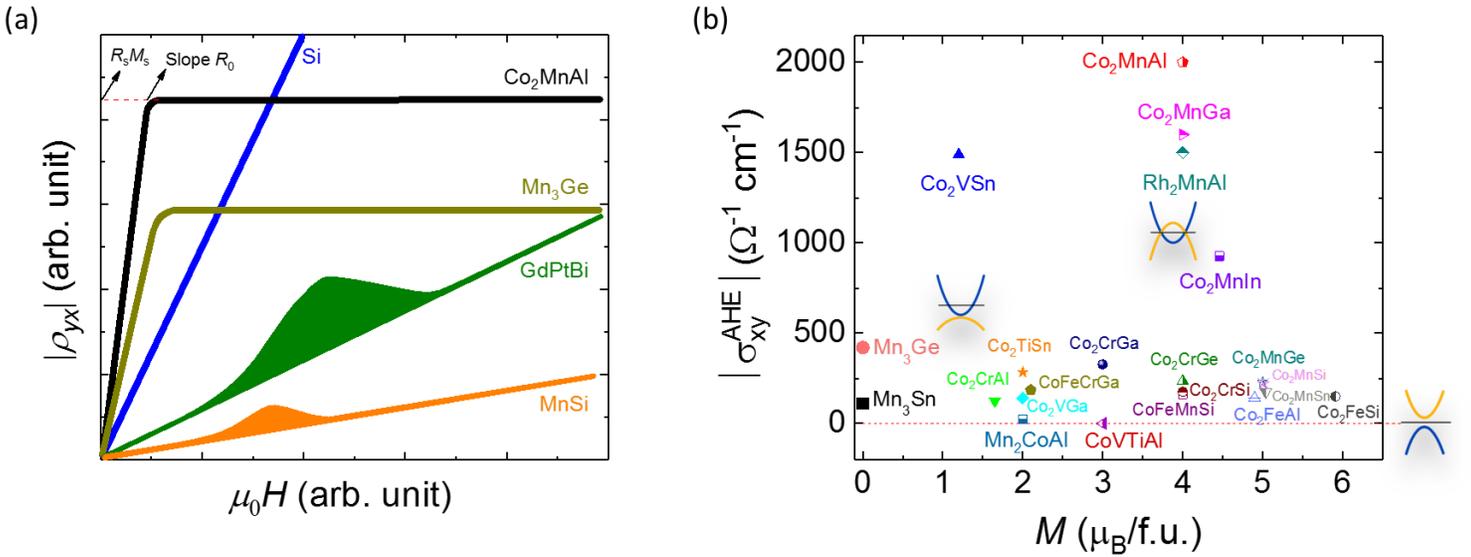

Fig. 5| **Tuning the anomalous Hall effect by Berry curvature.** (a) Nature of the Hall effect in various compounds related to this review. (b) Anomalous Hall conductivity vs magnetization of various Heusler compounds.

The AHE and THE are the two most important branches of Hall physics. Their use may provide a new direction to obtain novel spintronic devices for information storage and transfer and other applications. Owing to the tunability of magnetic and electronic structures, Heusler compounds not only provide numerous candidates for the realization of different Hall effects, but also offer new insights for the understanding of the fundamental physical principles involved in both effects. From a general understanding, in ferromagnetic or ferrimagnetic materials with finite magnetization, the Hall conductivity usually scales with sample magnetization[27]. However, from the large AHE in non-collinear antiferromagnets $Mn_3Ge$ and $Mn_3Sn$, we learn that the AHE can be observed in systems with vanishing net magnetic moments, where the only variable is the symmetry-dependent Berry curvature distribution, as shown in Fig. 5(a). A non-collinear spin texture due to skyrmion formation in MnSi gives rise to a distinct THE, which is known as the real-space Berry curvature effect[144]. Interestingly, this THE can also arise because of the momentum-space Berry curvature as in the AFM *half*-Heusler GdPtBi, where Weyl points form close to $E_F$ in an external magnetic field[25,72]. Plotting the AHC versus the magnetic moments of various Heusler compounds [Fig. 5(b)], one sees that AHC does not scale with sample magnetization and acquires a large value if the material possess a topological non-trivial band structure, such as nodal lines and Weyl points. Depending on how far the topological states reside from $E_F$ in the band structure, the transport characteristics modify accordingly.

From the topological point of view, it was already experimentally verified that Heusler compounds host a variety of non-trivial states such as TIs, WSMs, nodal line semimetals, magnetic skyrmions etc. Moreover, there are still many proposed topological states not yet verified, such as triple points[147], topological superconductors[34,148], Majorana fermions[36,149,150], Dirac semimetals[151] etc., which need further and extensive investigations. Heusler compounds provide a perfect platform to understand the interplay between the topology, crystal structure, and various physical properties beyond the conventional understanding[5,11]. With their highly tunable band structure, the Hall conductivity and charge carrier density can be adjusted to obtain giant anomalous Hall angles, which might lead to the realization of the long-expected quantum anomalous Hall effect at room temperature in thin Heusler films.

**Acknowledgements**

This work was financially supported by the ERC Advanced Grant No. 291472 `Idea Heusler', ERC Advanced Grant No 742068 – TOPMAT, and Deutsche Forschungsgemeinschaft DFG under SFB 1143.

Reprints and permissions information is available at www.nature.com/reprints.

**Competing financial interests**

The authors declare no competing financial interests.

**Correspondence**

Correspondence and requests for materials should be addressed to C. Felser (email: felser@cpfs.mpg.de).